\documentclass[conference]{IEEEtran}
\IEEEoverridecommandlockouts
\usepackage[bb=dsserif]{mathalpha}
\usepackage{cite}
\usepackage{amsmath,amssymb,amsfonts}
\usepackage{algorithmic}
\usepackage{graphicx}
\usepackage{textcomp}
\usepackage{xcolor}
\usepackage{graphicx, subfigure}
\usepackage{booktabs}
\usepackage{tabularx}
\usepackage{multirow}
\usepackage{xurl}
\usepackage{verbatimbox}
\usepackage{xcolor}
\usepackage{fancyvrb}
\usepackage{amsmath,amssymb}
\def\BibTeX{{\rm B\kern-.05em{\sc i\kern-.025em b}\kern-.08em
    T\kern-.1667em\lower.7ex\hbox{E}\kern-.125emX}}

\newcommand{\etal}{\textit{et al}., }

\begin{document}

\title{Memory Scraping Attack on Xilinx FPGAs:\\ Private Data Extraction from Terminated Processes
\thanks{This research was funded in part by a grant from the National Science Foundation.}
}


\author{\IEEEauthorblockN{Bharadwaj Madabhushi, Sandip Kundu, Daniel Holcomb}
\IEEEauthorblockA{\textit{Department of Electrical and Computer Engineering} \\
\textit{University of Massachusetts Amherst}\\
\{\textit{bmadabhushi, kundu, dholcomb}\}@umass.edu}
}
\maketitle

\begin{abstract}
FPGA-based hardware accelerators are becoming increasingly popular due to their versatility, customizability, energy efficiency, constant latency, and scalability.
FPGAs can be tailored to specific algorithms, enabling efficient hardware implementations that effectively leverage algorithm parallelism. This can lead to significant performance improvements over CPUs and GPUs, particularly for highly parallel applications. For example, a recent study found that Stratix 10 FPGAs can achieve up to 90\% of the performance of a TitanX Pascal GPU while consuming less than 50\% of the power. This makes FPGAs an attractive choice for accelerating machine learning (ML) workloads.
However, our research finds privacy and security vulnerabilities in existing Xilinx FPGA-based hardware acceleration solutions. These vulnerabilities arise from the lack of memory initialization and insufficient process isolation, which creates potential avenues for unauthorized access to private data used by processes.
To illustrate this issue, we conducted experiments using a Xilinx ZCU104 board running  the PetaLinux tool from Xilinx. We found that PetaLinux does not effectively clear memory locations associated with a terminated process, leaving them vulnerable to memory scraping attack (MSA). 
This paper makes two main contributions. The first contribution is an attack methodology of using the Xilinx debugger from a different user space. We find that we are able to access process IDs, virtual address spaces, and pagemaps of one user from a different user space because of lack of adequate process isolation. The second contribution is a methodology for characterizing terminated processes and accessing their private data. We illustrate this on Xilinx ML application library.

\end{abstract}


\section{Introduction}\label{section:Introduction}
\begin{figure}[h!]
  \centering 
    \includegraphics[width=0.42\textwidth]{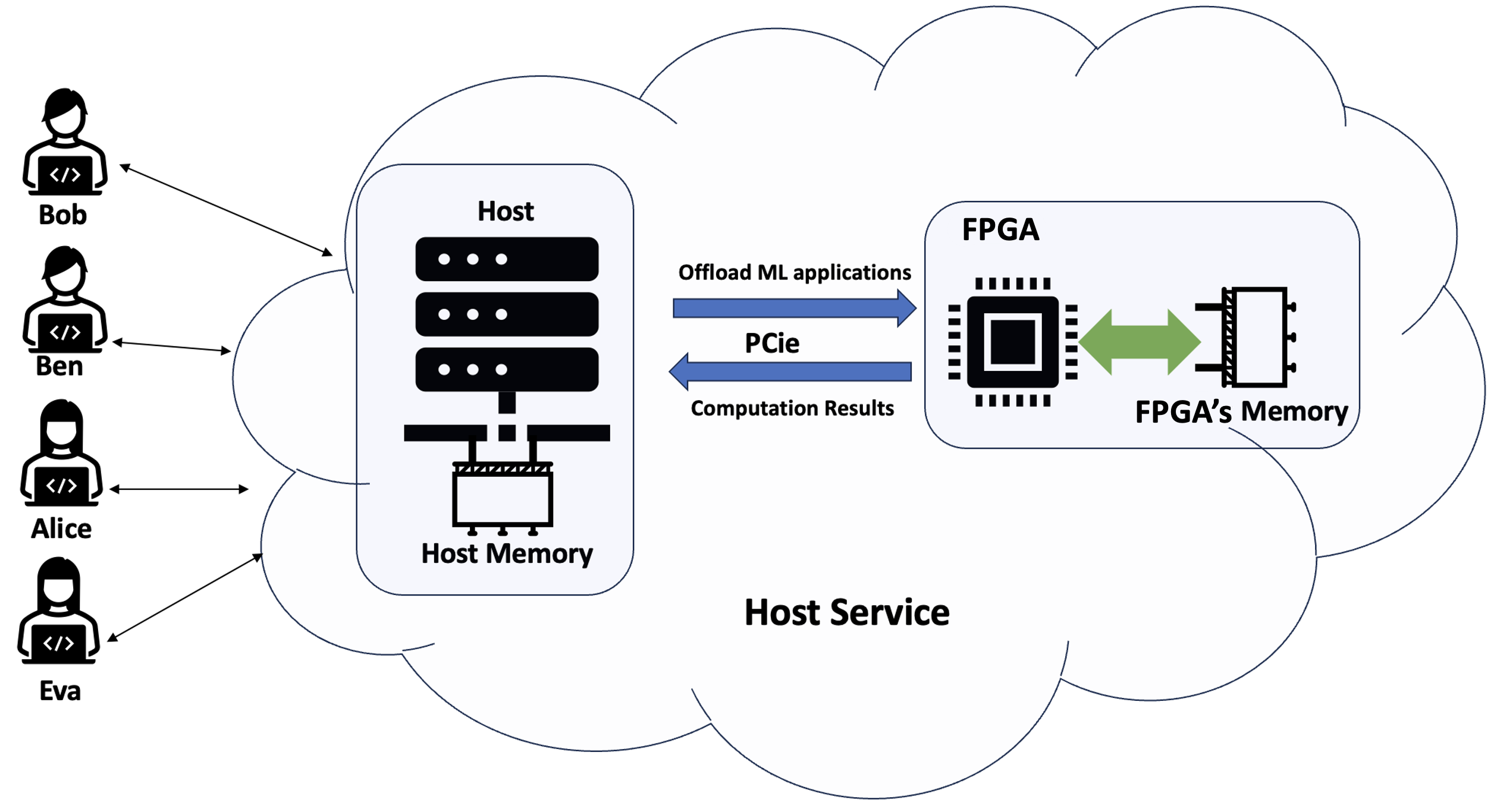} 
    \caption{A general host-based system.}
    \label{fig:i1}
\end{figure}

FPGAs have become increasingly popular as hardware accelerators in heterogeneous computing systems, especially in host-based environments such as data centers \cite{kachris2019hardware} and cloud computing systems \cite{7396187}. Their adoption is driven by their reconfigurable nature \cite{7577381}, lower power consumption \cite{hajirassouliha2018suitability}, high performance \cite{ullah2021high}, scalability \cite{6861604}, low cost \cite{putnam2014reconfigurable} and their ability to offload computationally intensive tasks from host CPUs \cite{sommer2017openmp}, thereby reducing the overall load on the host CPUs whilst providing better quality of service to the customers. Figure \ref{fig:i1} provides a general overview of the host-based system.

Multiple semiconductor companies are actively developing high-performance FPGAs to meet growing demand in heterogeneous computing systems, notably in data centers and for cloud service providers. AMD's Xilinx introduces Adaptive SOC FPGA \cite{asoc} and Zynq system-on-a-chip (SoC). Microsoft's Project Brainwave leverages FPGAs for real-time AI inference acceleration \cite{pbw}, prioritizing low latency and high performance in both cloud and edge environments. IBM incorporates cloud-FPGA \cite{ibm} solutions into its infrastructure as a service (IaaS), focusing on neural network modeling. Leading cloud providers, including AWS with Amazon EC2 F1 instances \cite{AWS}, leverage Xilinx  and Altera FPGAs  to offer FPGA acceleration services \cite{2022arXiv220911158K} commonly referring it as "FPGA-as-a-Service" (FaaS) or "Acceleration-as-a-Service" (AaaS)\cite{9581180}.

The integration of an FPGA into a host-based system introduces security concerns described earlier in \cite{9581180}.  However, the following risks were not considered earlier.

\subsection{Security Risks with Using Local Memory}

When offloading compute intensive tasks from host processing unit (PU) to FPGA for acceleration, the FPGA's local DRAM is used to temporarily store \cite{bobda2022future} and reuse data before returning results to the host. However, this poses a security risk, as a subsequent guest accessing the FPGA DRAM after the first guest's process has ended may be able to retrieve memory residue from the first user. We demonstrate this in this paper by scraping memory residue in FPGA DRAM from a terminated computer vision machine learning application.

\vspace{1ex}
In CPU, a memory management unit (MMU) enforces memory isolation between multiple processes\cite{mbongue2021deploying}. Similarly, in Xilinx FPGAs, a hypervisor like Xen manages isolation between multiple processes running on the FPGA \cite{hv}.  However, in CPUs, page tables are only accessible to the operating system (OS), not to any user process, including any program debugger a user may be running. We find that, unlike in CPUs, a Xilinx debugger has access to memory page tables. This is because Xen is not managed by the host OS, but rather configured by the user using PetaLinux (see Section \ref{section:background}). We find this to be a gaping security hole that affects not only Xilinx FPGAs but also other embedded systems. For the purpose of this paper, we limit our focus to FPGAs. 


\vspace{1ex}
\textit{Local Memory Scraping as an Attack Vector}: In this work, we show that (\textit{i}) Xilinx FPGAs do not perform automatic memory sanitization leaving memory residue, (\textit{ii}) Xilinx debugger can be invoked from a second user space (even for a single tenant FPGA), and  (\textit{iii}) page tables are accessible from the debugger. We present a memory scraping technique that uses the above exploits to show how sensitive information about previous programs can be reconstructed. 

\vspace{1ex}
\textit{Main Scientific Finding}: Our main scientific finding is that many accelerators, including FPGAs, GPUs and various embedded systems, have their own private memory that is not under direct host OS control. For performance reasons, it is not practical to have the host OS mediate every local memory access. This creates a general security problem, and in this paper, we demonstrate a targeted attack on Xilinx FPGAs to highlight this problem.

\subsection{Related Work}\label{section:related}
Zhou \etal \cite{zhou2016vulnerable}  and Maurice \etal \cite{maurice2014confidentiality} previously attacked NVIDIA's heterogeneous memory systems in cloud-based systems, allowing adversaries to access and reconstruct data from terminated processes. A similar issue affects ARM Mali GPUs, where a new process can access a terminated process's pages due to inadvertent reuse of freed pages \cite{gh}.


\vspace{1ex}
\textit{Memory Initialization Solutions}: A number of studies have investigated rapid DRAM initialization techniques. Seshadri \etal proposed \textit{RowClone}, a technique for initializing contiguous DRAM sections with zeros \cite{seshadri2013rowclone}. Seol \etal proposed a \textit{RowReset}, a hardware-efficient memory initialization solution that manipulates VDD and VSS to DRAM banks \cite{seol2017dram}. However, these methods are best suited for continuous memory locations. 
In virtual environments, Address Space Layout Randomization (ASLR) \cite{marco2014effectiveness} is employed to enhance security by randomizing DRAM memory locations, providing defense against memory corruption attacks. However, in multi-tenant FPGA settings with non-contiguous memory addresses \cite{seo2017sgx}, the aforementioned memory initialization solutions may inadvertently erase active guest user data. These solutions typically clear continuous memory locations when initializing memory for a terminated user, which can include active guest user data. Therefore, there is a need for more efficient solutions to initialize non-contiguous memory locations without jeopardizing active user data in local DRAM.

\subsection{Target Platform}\label{section:background}
Leading cloud service providers, such as Amazon EC2 F1 \cite{AMD} and Alibaba Cloud \cite{Alibaba}, have adopted AMD's Virtex Ultrascale+ FPGA family. Baidu's Apollo platform employs Zynq Ultrascale+ MPSoC FPGAs for self-driving vehicles \cite{Baidu}. All Ultrascale+ FPGAs \cite{ultrascale} have onboard (local) memory for use by offloaded processes onto the FPGA board.

For our attack demonstration, we target the Zynq Ultrascale+ MPSoC ZCU104 board because it is more affordable than high-end Ultrascale+ Virtex boards. The ZCU104 has an architectural design similar to other Ultrascale+ FPGAs, making our attack scenario credible and relevant. The ZCU104 incorporates essential components like quad-core ARM Cortex-A53 APU, dual-core Cortex-R5 RPU, Mali-400 MP2 GPU, a high-definition video codec, and programmable logic component fabricated using 16nm FinFET+ technology. For generalizability studies we have reverified the attack on Zynq Ultrascale+ MPSoC ZCU102 board.

\begin{figure*}[ht]
  \centering 
    \includegraphics[width=0.415\textwidth]{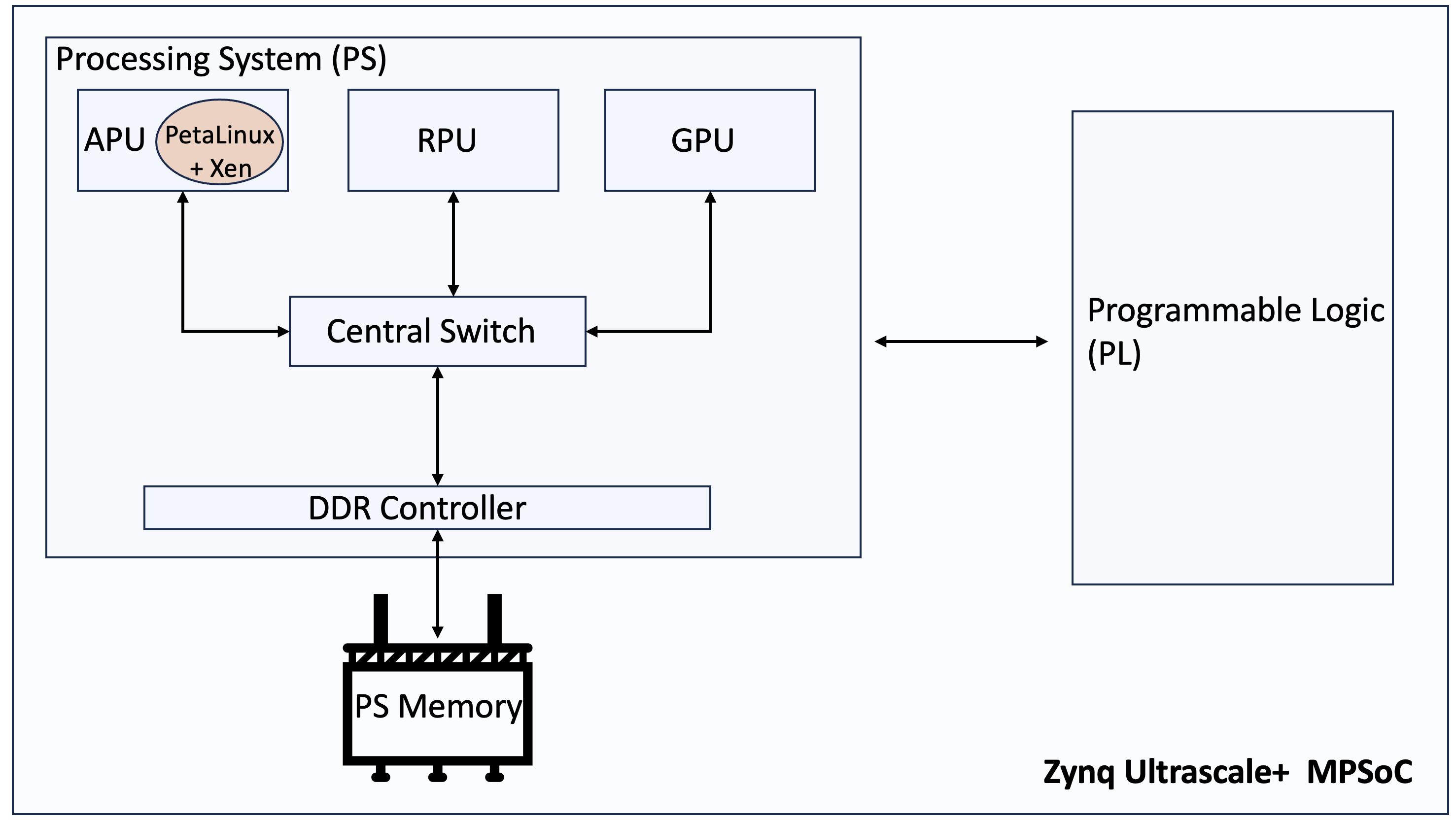} 
    \caption{A high level block diagram of Zynq Ultrascale+ MPSoC.}
    \label{fig:b2}
\end{figure*}

PetaLinux is the chosen software platform for ZCU104's system development \cite{Petalinux}. It offers tools like command-line interfaces, generators, templates, and system configuration tools, including the Xen hypervisor as a selectable component \cite{Xen}. PetaLinux's shell manager oversees hardware and software components for managing and functionality. Figure \ref{fig:b2} provides an architectural overview with PetaLinux running on the APU.

\subsection{Contributions}
The main contributions of this paper are as follows:

\begin{enumerate}
    \item We identify a security risk with using local memory.
    \vspace{1ex}
    \item We present a novel attack methodology that uses the Xilinx system debugger to mount a \textit{system-channel attack}. A debugger is typically used to inspect the values of variables and optimize code by identifying performance bottlenecks and memory leaks. It can also track and monitor specific memory locations. We exploit this feature, coupled with the fact that local memory is not controlled by the host OS, bypassing access control to mount our attack. Our attack works on both single-tenant and multi-tenant FPGAs.
    \vspace{1ex}
    \item We use our novel attack methodology to scrape memory of a terminated process. 
    \vspace{1ex}
    \item We present a novel data analysis technique involving offline profiling to learn high-value memory locations, then reading data from these locations in the scraped memory to reconstruct information about the target process.
    \vspace{1ex}
    \item We demonstrate our data analysis technique on Xilinx machine-learning model library identifying specific models used in terminated processes and revealing sensitive information such as input images and weights.
\end{enumerate}

\section{Adversary Model}\label{section:methodology}
This section outlines the adversary's goal, privileges, and capabilities.\vspace{1ex}

\noindent\textbf{Adversary's goal}: The adversary's goal is to (i) access data from local memory used previously by a terminated process and (ii) to perform analysis to reconstruct information from the terminated program. The adversary uses this information for compromising the privacy and security of the previous process entity. 

\vspace{1ex}
\noindent\textbf{Adversary's privileges}: Adversaries can use a system debugger provided by the manufacturer, resulting in having unrestricted access to critical process details, including process IDs (pids), virtual address spaces, and pagemaps. Normally, such privileges are not available under a CPU OS. However, Xilinx System Debugger permits these privileges from user space giving adversaries access to data stored in the FPGA's DRAM at physical locations associated with a specific process (pid).

\vspace{1ex}
\noindent\textbf{Adversary's access}: The adversary has access to FPGA libraries and IPs used provided by Xilinx that are used by the user. By profiling the FPGA libraries and IPs provided by Xilinx, the adversary can, for instance, gain an understanding of the physical memory layout of machine learning models running on the board. This allows the adversary to identify where critical data, such as weights, vectors, and images, are stored. During the attack, the adversary uses this profiling data to identify the model and locate critical data in the FPGA's DRAM. The adversary can then attempt to reconstruct the associated image, compromising the previous user's data confidentiality.

\vspace{1ex}
\section{Proposed Attack Methodology}
\label{sec:proposed-approach}
The adversary follows a four-step sequence to extract and analyze data from FPGA's DRAM, enabling them to identify the executed model and potentially reconstruct the image, highlighting their capabilities outlined in \ref{section:methodology}. The adversary's five steps are as follows:\vspace{1ex}

\begin{enumerate}
    \item \textbf{Polling for pid:} The adversary continuously monitors the system to identify the relevant process of interest, utilizing commands like "ps -ef" in Unix to extract the process ID (PID) associated with the targeted execution. \vspace{1ex}
    
    \item \textbf{Fetching virtual addresses and converting them to physical addresses:} Using the process ID, the adversary retrieves the virtual address locations of the targeted process from the heap mapping in the associated maps file. They then convert these virtual addresses into corresponding physical addresses within the FPGA's DRAM using information from the process's specific {\sf pagemaps} file. \vspace{1ex}
    
    \item \textbf{Data extraction from physical addresses:} Once the targeted process is terminated or disconnected, the adversary proceeds to access and read the contents of the previously derived physical address locations within the FPGA's DRAM. By doing so, they gain access to the data stored by the terminated process. \vspace{1ex}

    \item \textbf{Analysis of extracted data:} Once the data is extracted the adversary now proceeds to analyze the data. 
    \begin{enumerate}
        \item \textbf{\textit{Identifying models from strings:}} The adversary analyzes the FPGA DRAM data for distinct patterns or signatures of different models. Using criteria like keywords or known model names (e.g. "resnet50", "squeezenet"), they identify the model run by the targeted process based on the presence of similarly named libraries and data structures in memory. \vspace{1ex}
        
        \item \textbf{\textit{Reconstructing image:}} Depending on the model and whether it accepts an image as input, the adversary might attempt to reconstruct this input image. This is achievable due to the adversary's possession of knowledge about the physical memory layout of the identified model, which was acquired through offline profiling. Leveraging this information, the adversary can pinpoint the exact location where the image is stored and make an attempt at reconstruction. \vspace{1ex}
    \end{enumerate}
\end{enumerate}

\section{Experimental Setup}\label{section:experimentalsetup}
\begin{figure}[h!]
  \centering 
    \includegraphics[width=0.33\textwidth]{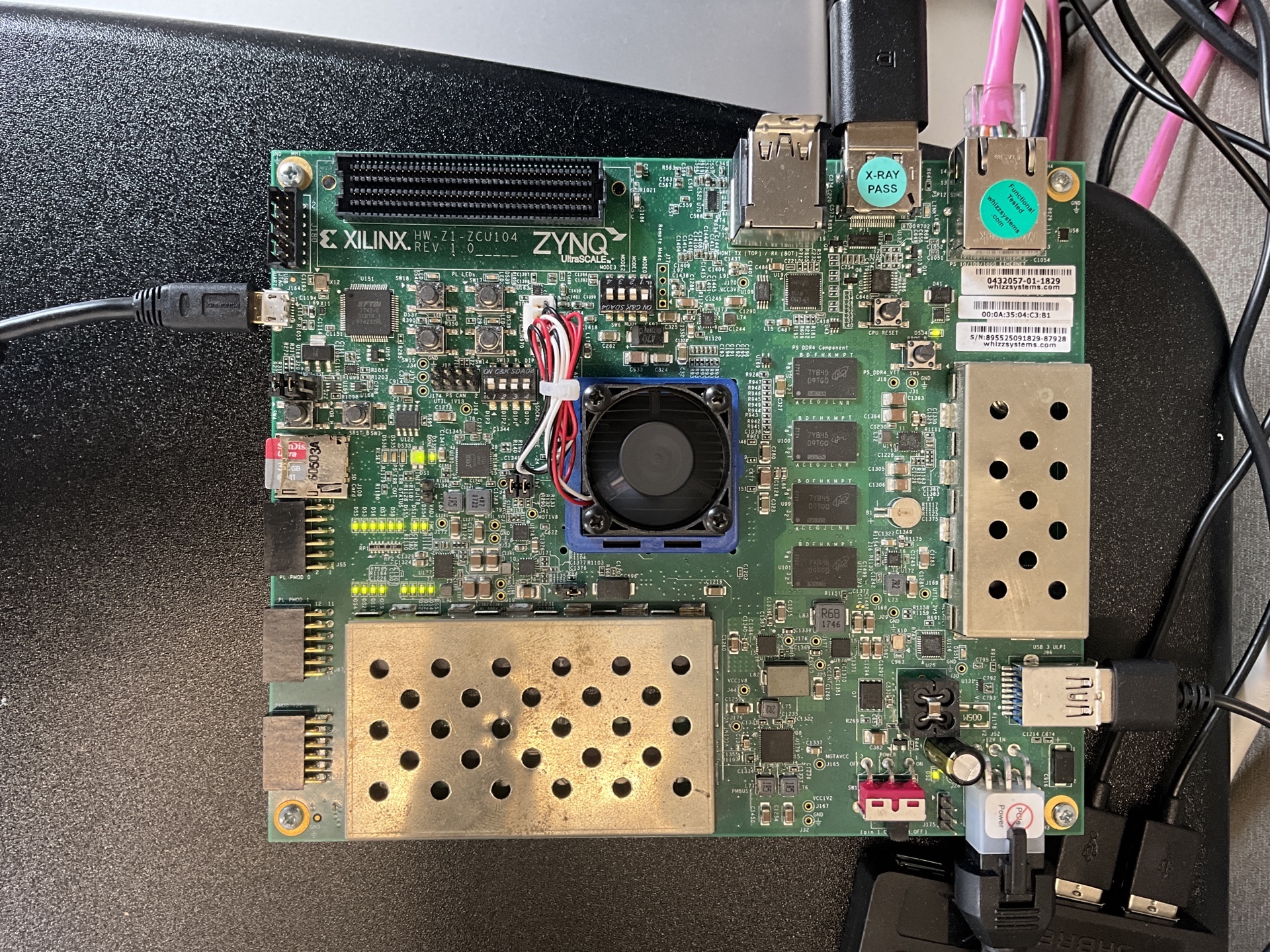} 
    \caption{Target Board (Xilinx's Zynq ZCU104)}
    \label{fig:zcu104}
\end{figure}

\noindent\textbf{Setting up Target board:} To conduct the experiments with the Xilinx ZCU104 FPGA board, we followed Xilinx's step-by-step instructions outlined in \cite{vitis}. Figure \ref{fig:zcu104} illustrates the target board.
\begin{enumerate}
    \item The Xilinx-provided OS image for the ZCU104 board is flashed to an SD card. This image contains the Petalinux embedded OS and necessary software tools. The SD card is then inserted into the ZCU104 board and powered on to boot the system.\vspace{1ex}
    \item After booting the board, we established a remote connection via the Ethernet interface to communicate and interact with it. \vspace{1ex}
    \item Finally, we installed the Vitis AI runtime on the target board, which provides various pre-built machine learning models from different vendors for testing and experimentation.
\end{enumerate}

\vspace{1ex}
\noindent\textbf{Victim model:} The selected victim model for our experimentation is {\sf "resnet50\_pt"} (RESidual NETwork using PyTorch Framework), sourced from Xilinx's examples. We chose this model because of its widespread use in image recognition tasks. Additionally, Xilinx has supplied a dedicated image designed for use with the resnet50\_pt model, which we employed in our experiments.

\begin{figure}[h!]
  \centering     
  \subfigure[Original image]{\label{fig:a}\includegraphics[width=0.30\textwidth]{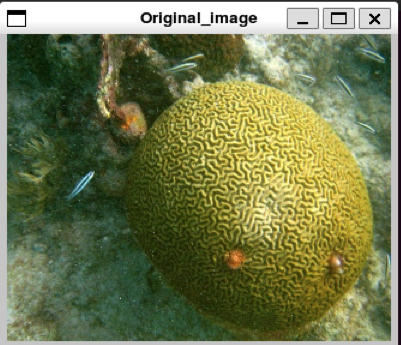}}
  \subfigure[Computed image]{\label{fig:b}\includegraphics[width=0.30\textwidth]{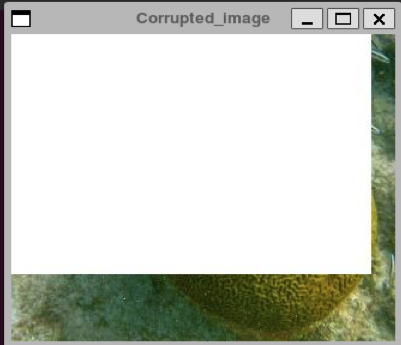}}
  \caption{The top image (a) represents an example input for the resnet50\_pt model, provided by Xilinx. The bottom image (b) shows a corrupted version achieved by altering specific pixel locations within the original image. About 20\% of the image has been intentionally omitted to highlight the original image is modified.}
  \label{fig:e1}
\end{figure}

\vspace{1ex}
\noindent\textbf{Corrupting the image:} We intentionally corrupted the example image by replacing its pixel values with 0xFFFFFF. When the adversary reads data from the FPGA's DRAM and encounters a sequence of 0xFFFFFF values, it signifies the corrupted image used as input for the resnet50\_pt model (as shown in Figure \ref{fig:e1}). This indicates that the corrupted image was not cleared from the FPGA's DRAM after the process terminated, underscoring the absence of proper memory management.

\vspace{1ex}
\noindent\textbf{Implementing steps described in Section \ref{sec:proposed-approach}:} The attack uses two terminals: one for the attacker and one for the victim. The victim runs the resnet50\_pt model while the attacker runs Steps 1 and 2. After the victim's process ID disappears, confirming it has ended, the attacker proceeds with Steps 3 and 4 in the attacker terminal to read the data from the FPGA's DRAM and identify the executed model and try to reconstruct images. Our code written in python automates the full attack process.

\section{Results}\label{section:results}
In this section, we provide  results with illustrations from each step described in Section \ref{sec:proposed-approach}. These results also illustrate the implementation aspects of the attack.

\vspace{1ex}
\noindent\textbf{Step 1. Polling for pids}: Figures \ref{fig:r1} and \ref{fig:r2} illustrates the running processes (pids) obtained from the attacker's terminal by executing the {\sf ps -ef} command. Figure \ref{fig:r1} shows the processes before running the {\sf resnet50\_pt} model, while Figure \ref{fig:r2} shows the processes after its execution.
\begin{figure}[h!]
\begin{verbbox}[\scriptsize\slshape]
1389       2  0 03:51 ?        00:00:00 [kworker/3:0-events]
1390     843  0 03:52 pts/0    00:00:00 ps -ef
\end{verbbox}
\resizebox{8.6cm}{0.7cm}{\fbox{\theverbbox}}
\caption{(Step 1) Process list before victim model was run.}
\label{fig:r1}
\end{figure}
\begin{figure}[h!]
\begin{verbbox}[\scriptsize\slshape]
1389       2  0 03:51 ?        00:00:00 [kworker/3:0-events]
1391    2430 18 12:33 pts/1    00:00:00 ./resnet50_pt 
                                        /usr/share/vitis_ai_library/models/resnet50_pt/resnet50_pt.xmodel 
                                        ../images/001.jpg
1392    1875  0 12:33 pts/0    00:00:00 ps -ef
\end{verbbox}
\resizebox{8.6cm}{2.0cm}{\fbox{\theverbbox}}
\caption{(Step 1) Process list after Victim model was run. Victim's pid is observed to be 1391.}
\label{fig:r2}
\end{figure}

\vspace{1ex}
\noindent\textbf{Step 2. Fetching virtual addresses and converting them to physical addresses:} We access the process's memory map using the command {\sf vim /proc/1391/maps} for {\sf PID 1391}, revealing the virtual address range of the heap, from {\sf 0xaaaaee775000} to {\sf 0xaaaaefd8a000} as shown in Figure \ref{fig:r3}. To convert these heap virtual addresses to physical addresses in the FPGA's DRAM, we have created C code based on our offline training knowledge, as explained in Section \ref{section:methodology}. This code maps the virtual addresses to a physical range, visible in Figure \ref{fig:r4}, from {\sf 0x61c6d730} to {\sf 0x61ec5e220}.
\begin{figure}[h!]
\begin{verbbox}[\scriptsize\slshape]
aaaaee775000-aaaaefd8a000 rw-p 00000000 00:00 0                          [heap]
ffffb13b5000-ffffb6c1f000 rw-p 00000000 00:00 0             /dev/dri/renderD128
\end{verbbox}
\resizebox{8.6cm}{0.7cm}{\fbox{\theverbbox}}
\caption{(Step 2) Virtual address of the target process ranges from 0xaaaaee775000 to 0xaaaaefd8a000 in the heap.}
\label{fig:r3}
\end{figure}

\begin{figure}[h!]
\begin{verbbox}[\scriptsize\slshape]
xilinx-zcu104-20222:~# ./virtual_to_physical.out 1391 0xaaaaee775000
0x61c6d730
xilinx-zcu104-20222:~# ./virtual_to_physical.out 1391 0xaaaaefd8a000
0x61ec5e220
\end{verbbox}
\resizebox{8.6cm}{1.3cm}{\fbox{\theverbbox}}
\caption{(Step 2) Physical address values of the virtual addresses.}
\label{fig:r4}
\end{figure}

\vspace{1ex}
\noindent\textbf{Step 3. Data extraction from physical addresses:} To continuously monitor the termination of the specified process ID (PID), we repeatedly execute step 2. If the PID has been terminated successfully, it will no longer appear in the list of running processes. Figure \ref{fig:r5} illustrates the absence of the PID from the process running list after its termination. 
\begin{figure}[h!]
\begin{verbbox}[\scriptsize\slshape]
1389       2  0 03:51 ?        00:00:00 [kworker/3:0-events]
1401    1875  0 12:33 pts/0    00:00:00 ps -ef
\end{verbbox}
\resizebox{8.6cm}{0.5cm}{\fbox{\theverbbox}}
\caption{(Step 3) The figure shows that the PID 1391 absent in the process running list after it was terminated}
\label{fig:r5}
\end{figure}

\begin{figure}[h!]
\begin{verbbox}[\scriptsize\slshape]
xilinx-zcu104-20222:~# devmem 0x61c6d730
0x00000000
xilinx-zcu104-20222:~# devmem 0x61ec5e220
0xF7F5F8FD
\end{verbbox}
\resizebox{8.6cm}{1.2cm}{\fbox{\theverbbox}}
\caption{(Step 3) The figure provides an example of how "devmem" is utilized to read the data.}
\label{fig:r6}
\end{figure}

We proceeded by executing the command {\sf "devmem \textlangle physical\_address\textrangle" } to retrieve data from the physical address locations of the FPGA's DRAM obtained in Step 3. Figure \ref{fig:r6} shows an example of how devmem command is used to read the data. However, since these steps are automated, the {\sf devmem} command is executed for all the physical address locations specified in Step 2.

\vspace{1ex}
\noindent\textbf{Step 4.a Analysis of extracted data (Identifying models from strings):} After entire data is extracted in Step 4, we now format this data it into a file, arranging the data into rows of eight nibbles each. Subsequently, we create a hex dump of this file by running {\sf "hexdump" } on it to inspect if any meaningful, readable words emerge. By analyzing the snippet displayed in Figure \ref{fig:r7}, we discern that the converted string representation of the hex data reveals the presence of the model name {\sf resnet50\_pt} in the data readout.
\begin{figure}
\begin{verbbox}[\scriptsize\slshape]
xilinx-zcu104-20222:~# grep "resnet50" 1391_hexdump.log
6c73 2f72 6573 6e65 7435 305f 7074 2f72  ls/resnet50_pt/r
6876 6973 696f 6e2f 7265 736e 6574 3530  hvision/resnet50
\end{verbbox}
\resizebox{8.5cm}{!}{\fbox{\theverbbox}}
\caption{(Step 4.a) The figure shows that the model resnet50\_pt is found in the read out from the FPGA memory.}
\label{fig:r7}
\end{figure}

\vspace{1ex}
\noindent\textbf{Step 4.b Analysis of extracted data (Reconstructing image):} Once it is established that the executed model corresponds to "resnet50\_pt," we proceed with image reconstruction by searching for the identifier {\sf "FFFF FFFF"} in the hexdump log, which signifies the corrupted image used by the identified model. Figure \ref{fig:r8} demonstrates the hexdump file created in Step 4.a and the observation of the image identifier in this file. This highlights that the data associated with pid 1391 is not cleared from the DRAM even after termination.

In practical experiments, we varied the pixel values of the input image. To precisely locate the image's starting point within the hexdump, we conducted offline profiling by changing pixel values to "0x555555." We then ran the "resnet50\_pt" model offline with this modified image, repeating Steps 1 to 3. By analyzing the hexadecimal dump, we found the offset between the first occurrence of "5555 5555" and the hexdump file's start, specifically at row number "646768." As we only modified the image, preserving the underlying model's integrity, the image's offset within the heap remained consistent for any image used with this model. Utilizing this profiled information, we successfully retrieved and reconstructed the victim's input image from the data at the identified offset position within the victim's heap file, which was previously saved. This streamlined process enabled the reconstruction of the input image, leveraging insights gained from profiling various existing models.
\begin{figure}[h!]
\begin{verbbox}[\scriptsize\slshape]
0000 0000 0000 0000 9102 0000 0000 0000  ................
8007 71f1 aaaa 0000 7012 71f1 aaaa 0000  ..q.....p.q.....
....
....
FFFF FFFF FFFF FFFF FFFF FFFF FFFF FFFF  UUUUUUUUUUUUUUUU
FFFF FFFF FFFF FFFF FFFF FFFF FFFF FFFF  UUUUUUUUUUUUUUUU
FFFF FFFF FFFF FFFF FFFF FFFF FFFF FFFF  UUUUUUUUUUUUUUUU
FFFF FFFF FFFF FFFF FFFF FFFF FFFF FFFF  UUUUUUUUUUUUUUUU
FFFF FFFF FFFF FFFF FFFF FFFF FFFF FFFF  UUUUUUUUUUUUUUUU
....
....
FFFF FFFF FFFF FFFF FFFF FFFF FFFF FFFF  UUUUUUUUUUUUUUUU
FFFF FFFF FFFF FFFF FFFF FFFF FFFF FFFF  UUUUUUUUUUUUUUUU
FFFF FFFF FFFF FFFF FFFF FFFF FFFF FFFF  UUUUUUUUUUUUUUUU
FFFF FFFF FFFF FFFF FFFF FFFF FFFF FFFF  UUUUUUUUUUUUUUUU
FFFF FFFF FFFF FFFF FFFF FFFF FFFF FFFF  UUUUUUUUUUUUUUUU
...

\end{verbbox}
\resizebox{8.6cm}{5.0cm}{\fbox{\theverbbox}}
\caption{(Step 4.b) The figure shows the occurrence of "FFFF FFFF" the identifier of corrupted image used as an input by the model.}
\label{fig:r8}
\end{figure}

\section{Conclusion and Future Work}\label{section:conclusion}
In this paper, we identified a security gap with using local memory that is not under host OS control. Typically, FPGA (or similar accelerator) manufacturers like to manage their own local memory for performance and efficiency. However, they must also equip their users to debug their program giving them access to local memory content from the debugger. Since the debugger accesses the local accelerator memory without host OS mediation, it falls on the FPGA manufacturer to restrict debugger access privileges. In this context, we find that the PetaLinux tool used by Xilinx to manage an FPGA has major security holes. First, it allows unrestricted access to the page map tables. This enables an attacker process to scrape memory from a terminated victim process. Second, when a process terminates, it does not sanitize the physical memory used by the terminated process. Thus, an attacker can access the memory pages used by a terminated process. Third, it does not use any kind of randomization in physical page layout. This allows an attacker to learn about input or output data offsets, simply by learning from running the same program with its own input data. PetaLinux is a Xilinx supported tool to manage its FPGA cards. We conducted experiments on the Xilinx ZCU104 board using the Xilinx SDK to execute a machine learning program (referred to as the victim process). Through a systematic and step-by-step approach, we successfully showcased how an attacker can gain access to memory pages from the process, allowing them to deduce the specific program that was running and discern the input used. To enhance reproducibility and enable further exploration, we have automated the attack process and plan to release our code on GitHub.

\section{Ethical Disclosure}\label{ethical}
In line with responsible disclosure and ethical practices within the computer security research community, we reported these findings to AMD/Xilinx on July 14, 2023, along with all relevant details. AMD, the parent company of Xilinx acknowledged the validity of the attack on August 23, 2023.  

\bibliographystyle{IEEEtran}
\bibliography{crypto}
\end{document}